# ARPES studies of the electronic structure of LaOFe(P,As)

D. H. Lu[1*], M. Yi[2,3], S.-K. Mo[3,4], J. G. Analytis[2,3], J.-H. Chu[2,3], A. S. Erickson[2,3], D. J. Singh[5], Z. Hussain[4], T. H. Geballe[2,3], I. R. Fisher[2,3], and Z.-X. Shen[2,3]

[1]Stanford Synchrotron Radiation Lightsource, SLAC National Accelerator Laboratory, 2575 Sand Hill Road, Menlo Park, CA 94025.

[2]Stanford Institute for Materials and Energy Sciences, SLAC National Accelerator Laboratory, 2575 Sand Hill Road, Menlo Park, CA 94025.

[3]Geballe Laboratory for Advanced Materials, Departments of Physics and Applied Physics, Stanford University, CA 94305.

[4]Advanced Light Source, Lawrence Berkeley National Lab, Berkeley, CA 94720, USA

[5]Materials Science and Technology Division, Oak Ridge National Laboratory, Oak Ridge, Tennessee 37831-6114, USA

**ABSTRACT**

We report a comparison study of LaOFeP and LaOFeAs, two parent compounds of recently discovered iron-pnictide superconductors, using angle-resolved photoemission spectroscopy. Both systems exhibit some common features that are very different from well-studied cuprates. In addition, important differences have also been observed between these two ferrooxypnictides. For LaOFeP, quantitative agreement can be found between our photoemission data and the LDA band structure calculations, suggesting that a weak coupling approach based on an itinerant ground state may be more appropriate for understanding this new superconducting compound. In contrast, the agreement between LDA calculations and experiments in LaOFeAs is relatively poor, as highlighted by the unexpected Fermi surface topology around $(\pi,\pi)$. Further investigations are required for a comprehensive understanding of the electronic structure of LaOFeAs and related compounds.

* Corresponding author. 2575 Sand Hill Road, Menlo Park, CA 94025, USA. Tel.: +1 650 926 3026; fax: +1 650 926 4100; E-mail address: dhlu@slac.stanford.edu.

# 1. Introduction

The recent discovery of superconductivity in iron-based layered compounds has created renewed interest in high temperature superconductivity [1-9]. With a superconducting transition temperature as high as 55 K [7], this discovery provides a new playground to understand the essential ingredients for achieving a high superconducting transition temperature. Some early experiments seem to hint at a strong similarity with the cuprate superconductors, such as the close proximity to a magnetically ordered parent phase [10-12]. Extensive theoretical investigations have been carried out on the mechanism. A burning current issue is the nature of the ground state of the parent compounds, and two distinct classes of theories have been put forward characterized by contrasting underlying band structures: local moment antiferromagnetic ground state for strong coupling approach [13-17] and itinerant ground state for weak coupling approach [18-22]. The local moment magnetism approach stresses on-site correlations and proximity to a Mott insulating state and thus a resemblance to cuprates; while the latter approach emphasizes itinerant electron physics and the interplay between competing ferromagnetic and antiferromagnetic fluctuations.

Such a controversy is partly due to the lack of conclusive experimental information on the electronic structures. Here we report the angle-resolved photoemission spectroscopy (ARPES) investigation of LaOFeP and LaOFeAs, the parent compounds of iron-pnictide superconductors [2, 3]. Our data reveal a number of common features between these two parent compounds that clearly distinguish them from parent compounds of cuprate superconductors: i) they have much higher density of states near the Fermi level; ii) they have multiple bands and Fermi surface sheets; iii) there is no evidence of the pseudogap effect that is so pervasive in underdoped cuprates. Furthermore, our results unveil some important differences between these two oxypnictides although the LDA band structure calculations for these two compounds appear to be very similar. For LaOFeP, our data exhibit remarkable agreement with LDA band structure calculations, therefore strongly favoring the itinerant ground state [23]. For LaOFeAs, the agreement between our data and the calculations is relatively poor. The exact origin of such a disparity is unclear at this point and deserves further investigations.

## 2. Experiment

Single crystals of LaOFeP and LaOFeAs, with dimensions up to 0.4 × 0.4 × 0.04 mm$^3$, were grown from a tin flux, using modified conditions from those first described by Zimmer and coworkers [24], as described elsewhere [25]. While the undoped LaOFeAs is not superconducting, the superconducting transition temperature ($T_c$) of LaOFeP, determined from resistivity ($\rho$) and susceptibility measurements, was 5.9±0.3 K. Residual resistance ratios, $\rho(300K)/\rho_0$, were up to 85 for LaOFeP, indicative of high crystal quality. ARPES measurements were carried out at beamline 10.0.1 of the Advanced Light Source (ALS) using a SCIENTA R4000 electron analyzer. All data presented in this paper were recorded using 42.5 eV photons. The total energy resolution was set to 16 meV and the angular resolution was 0.3°. Single crystals were cleaved *in situ* and measured at 20 K in an ultra high vacuum chamber with a base pressure better than 2×10$^{-11}$ Torr. Electronic structure calculations were performed within the local density approximation using the general potential linearized augmented planewave (LAPW) method. The convergence of the basis set and zone sampling was checked. Local orbitals were used to relax linearization errors and to include the semicore levels of the metal atoms. The calculations shown in this paper were done using the experimental lattice parameters but with relaxed internal atomic positions. There is a sensitivity of the Fermi surface to these coordinates, and as a result the present Fermi surface differs somewhat from that of Lebegue [22] who did not relax these coordinates.

## 3. Results and discussions

Fig. 1 compares the angle-integrated photoemission spectrum (AIPES) with the density of states (DOS) obtained from the LDA band structure calculations for LaOFeP and LaOFeAs. Both AIPES spectra consist of a sharp intense peak near the Fermi level ($E_F$) that is separated from the main valence band (VB) peaks at higher binding energy. According to the LDA calculations, the near-$E_F$ states have dominant Fe *d* character while the peaks at higher binding energy are mixtures of O *p* states and hybridized Fe *d* and pnictogen *p* states. Compared with the calculated DOS, the near-$E_F$ peak in both materials has a narrower width than the calculated Fe *d* states and is pushed closer to $E_F$, which is consistent with the band renormalization effect as we will discuss later. The VB peaks at higher binding energy, on the other hand, are shifted towards higher binding energy, resulting in slightly larger total VB width. It is important to note that such a VB spectrum with an intense peak near $E_F$ is in sharp contrast with the typical VB spectrum of cuprates as shown in the inset of Fig. 1a. The VB spectrum of cuprates is characterized by a weak feature near $E_F$ on top of a broad VB peak, consistent with the doped Mott insulator picture. This clear disparity between the iron pnictides and cuprates suggests that itinerant rather than Mott physics is a more appropriate starting point for the iron-based superconductors. We also notice some subtle differences between LaOFeP and LaOFeAs. Although the near-$E_F$ peak is strong in both cases, it appears to be less intense in LaOFeAs, almost a factor of 2 smaller than in LaOFeP with respect to the main VB peak. It is unclear at this point whether it is due to stronger correlation effect in LaOFeAs, or some subtle band structure effect. In addition, the multiple peak structure in the VB is less pronounced in LaOFeAs, which may reflect different level of hybridization between Fe $d$ and pnictogen $p$ states in these two compounds.

More detailed information can be obtained from angle-resolved data. Fig. 2 shows the Fermi surface (FS) mapping results of these two compounds. For LaOFeP (Fig. 2a&2b), three sheets of Fermi surfaces were clearly observed: two hole pockets ($\Gamma_1$ and $\Gamma_2$) centered at Γ and one electron pocket (*M*) centered at M. Note that five sheets of Fermi surfaces were predicted in band structure calculations: two hole pockets around Γ, two electron pockets around M, and one heavily 3D hole pocket centered at Z [22]. As we will show later, the inner hole pocket $\Gamma_1$ observed in our data should contain two nearly degenerate sheets, same for the electron pocket around M. Therefore, the observed FS topology is in good agreement with the LDA calculations in terms of the number of sheets and the character of each sheet (hole *vs.* electron). For LaOFeAs (Fig. 2c&2d), again two hole pockets ($\Gamma_1$ and $\Gamma_2$) centered at Γ can be observed, which are very similar to those in LaOFeP. In addition, there is a small bright patch right at Γ. A close examination of the data taken along the high symmetry cut reveals that this patch originates from an intense hole-like band at Γ with its band top almost touching the Fermi level. Therefore, it is not a true Fermi surface sheet. Surprisingly, a cross-shaped Fermi surface centered at M is observed, which looks very different from the small electron pocket observed in LaOFeP and is totally unexpected from the LDA calculations.

With the FS topology in mind, let us take a look at the ARPES spectra taken along the high symmetry lines. Fig. 3a and 3b present the data from LaOFeP along Γ-X and Γ-M, respectively. To understand the seemingly complex multi-band electronic structure, we superimpose the LDA band structure calculations on top of our data. A quantitative agreement can be found between the ARPES spectra and the calculated band dispersions after shifting the calculated bands up by ~0.11 eV and then renormalizing by a factor of 2.2. Along the Γ-X direction (Fig. 3a), two bands crossing $E_F$ can be clearly identified: one near the Γ-point ($\Gamma_1$) and one near the X-point ($\Gamma_2$), corresponding to the two hole pockets centered at Γ shown in Fig. 2b. According to the LDA calculations, the inner pocket originates from Fe $d_{xz}$ and $d_{yz}$ bands that are degenerate at Γ, and the splitting of these two bands close to Γ is too small to be resolved in our data. However, we do see evidence for the splitting at higher binding energy. The outer pocket is derived from the Fe $d_{3z^2-r^2}$ states that hybridize with the P $p$ and La orbitals. This band has strong $k_z$ dispersion and is very sensitive to the level of hybridization. Along Γ-M direction (Fig. 3b), in total three $E_F$ crossings can be resolved. In addition to the two crossings associated with the two hole pockets discussed above, another crossing near the M-point can be observed, although the corresponding crossing in the 2$^{nd}$ zone is too weak to be seen due to the matrix element effect. This crossing is responsible for the electron pocket centered at M. The LDA calculations also predict two bands crossing $E_F$ around the M-point, which cannot be clearly resolved in our data.

We should point out that a factor of 2.2 in band width renormalization does not necessary imply strong correlation effect in this compound. This value is actually comparable to that of $Sr_2RuO_4$, which is a correlated Fermi liquid and is reasonably well described by theories using itinerant band structure as the starting point [26]. Note that the values of the $E_F$ shift and band renormalization factor in Fig. 2 are chosen to obtain the best match of the two higher binding energy bands at the Γ-point. While the renormalized bands using this set of parameters fit the $\Gamma_1$ band very well, the match near the X-point and the M-point is less perfect. This suggests that different bands may have slightly different renormalization effects. Nevertheless, the overall degree of agreement between the experiments and the calculations is rather remarkable, indicating that the LDA calculations assuming an itinerant ground state capture the essence of the electronic structures of this system.

Such a quantitative agreement cannot be found in the high symmetry cuts of LaOFeAs (Fig. 3c&3d). Along the Γ-X direction (Fig. 3c), three hole-like bands centered at Γ can be clearly resolved. Among them, the band that crosses $E_F$ is associated with the $\varGamma_1$-pocket shown in the FS mapping. Note that the band associated with the $\varGamma_2$-pocket is completely suppressed in this geometry due to the matrix element effect, but is clearly visible in the cut perpendicular to this direction (not shown). These two $E_F$ crossing bands are very similar to those in LaOFeP, therefore are consistent with the LDA calculations. The other two bands that do not cross $E_F$ (the top one almost touches $E_F$ at Γ) are not seen in LaOFeP and their origin remains unclear at this point. More discrepancy with the LDA calculations lies in the bands near the M-point as shown in Fig. 3d. In comparison with the same cut in LaOFeP, it appears that the electron-like band and the flat hole-like band are pushed up and the bottom of the electron band is now located at ~0.05 eV compared with ~0.11 eV in LaOFeP. In addition, another hole-like dispersion very close to $E_F$, along with the electron-like band, are responsible for the cross-shaped FS sheet centered at the M-point. Due to the extremely weak intensity of features around the M-point, we cannot provide a satisfactory identification for each individual band as in LaOFeP. Further systematic investigations, including momentum, temperature and doping dependence studies, are required to achieve a comprehensive picture of the electronic structure in LaOFeAs.

It is worthwhile to mention that the similar band dispersions and the unexpected FS topology around the zone corner (π,π) have also been observed recently in the $Ba_{1-x}K_xFe_2As_2$ system [27]. With a careful comparison with the LDA calculations, the experimental data may be reasonably well described by the calculations after some shift and renormalization of the calculated bands [28]. We should also point out that there is an important difference in the ground state properties between LaOFeP and LaOFeAs even though the LDA calculations for these two systems appear to be very similar. It has been well established from neutron scattering experiments that LaOFeAs has a collinear antiferromagnetic ground state with a small ordered moment [10]. In contrast, no such a magnetic ground state has been reported for LaOFeP. Therefore, one would expect that the low temperature electronic structure would be somewhat different between these two compounds. Indeed, some band splitting associated with the SDW transition has been reported for both $BaFe_2As_2$ and $SrFe_2As_2$ systems [29, 30], and will be discussed in details in a separate paper. In this regard, it is not surprising to see that a nonmagnetic band structure calculation cannot fully describe the experimental results. A more sophisticated calculation taking into account the magnetic ordering, spin orbit coupling, and orthorhombicity may be needed to better match the experimental data and thereby improve our understanding of the electronic structure in these important parent compounds.

Fig. 4 displays the energy-distribution-curves (EDC's) along the same high symmetry cuts as shown in Fig. 3. Taking a close look at these EDC's, we find that there is no evident pseudogap in all bands crossing the Fermi level, in contrast to the ubiquitous observation of pseudogap in underdoped cuprates [31]. The absence of the pseudogap in both LaOFeP and LaOFeAs, therefore, marks another important difference between this new iron-based superconductor and cuprates. This finding also contradicts some early reports of a 20 meV pseudogap in both LaOFeP and LaOFeAs from AIPES [32]. The difference can be attributed to either a poor surface quality of the polycrystalline samples used for that measurement, where previous experience indicates potential problems associated with impurities [33], or distortion of the AIPES result by states away from $k_F$. ARPES from single crystalline samples is much better suited to address the pseudogap issue by directly measuring the states near $k_F$.

Finally, we shall comment on the FS volume counting issue. For LaOFeP, the FS volume enclosed under three pockets yields 1.94, 1.03, and 0.05 electrons for $\Gamma_1$, $\Gamma_2$ and $M$ pockets, respectively. Taking into account the unresolved, nearly degenerate sheets under $\Gamma_1$ and $M$ pockets, a total electron count of $5.0 \pm 0.1$ is obtained, which is roughly one electron less than the expected value of 6. This is in fact consistent with the fact that we need to shift $E_F$ of the calculated band structures down in order to match the measured band dispersions in Fig. 3a and 3b. In this regard, a recent de Haas-van Alphen (dHvA) effect measurement on LaOFeP reported a slightly different FS topology with a total electron count very close to the nominal value [34]. The major discrepancy between these two measurements lies in the volume enclosed by the $\Gamma_2$ pocket, which clearly indicates a subtle difference between bulk and surface. For LaOFeAs, the corresponding values are 1.86 and 1.18 electrons for $\Gamma_1$ and $\Gamma_2$ pockets, respectively. It is not easy to determine the exact volume under the *M* pocket as the cross-shaped FS is more like FS patches with no well defined $k_F$. On the other hand, the volume under this pocket is very small in any case so that the uncertainty is negligible. Therefore, the total electron count is again close to 5.0 as in LaOFeP (under the same assumption of a degenerate $\Gamma_1$ pocket). It is interesting to note that the same problem exists in a recent ARPES measurement of $NdFeAsO_{0.9}F_{0.1}$, an electron-doped superconductor [35]. The reported FS consists of a large hole pocket at $\Gamma$ and a small electron pocket around M, suggesting that the surface is in fact hole-doped. All these discrepancies seem to suggest a substantial change in the doping level on the cleaved surface, which is likely due to the lack of a charge neutral cleavage plane. If so, it would be a common problem for all 1111 family of iron oxypnictides. Despite this disagreement, the fact that all the expected Fermi surface pieces are observed in good agreement with respect to their Brillouin zone locations and signs (hole *vs.* electron) and the agreement in dispersion to great details as shown in Fig. 2 makes a strong case that the itinerant band structure captures the essence of the electronic structure of LaOFeP.

## 4. Conclusion

We present a comparison study of two parent compounds of iron pnictides, LaOFeP and LaOFeAs. It is clear from our data that both compounds share a number of common features that are very different from parent compounds of cuprates: i) they have much higher density of states near the Fermi level; ii) they have multiple bands and Fermi surface sheets with both hole pockets at $\Gamma$ and electron pockets at M; iii) there is no evidence of the pseudogap effect that is so pervasive in underdoped cuprates. On the other hand, our data also reveal some important differences between these two compounds in comparison with the LDA band structure calculations. For LaOFeP, our data exhibit remarkable agreement with LDA band structure calculations, therefore strongly favor the itinerant ground state. For LaOFeAs, the agreement between our data and the calculations is relatively poor, suggesting that a nonmagnetic LDA calculation is insufficient to describe the ground state properties in LaOFeAs and related FeAs systems. The origin of such a disparity between these two sister compounds is of great importance to understanding the physics in FeAs superconductors and deserves further systematic investigations.

**Acknowledgements**

We thank C. Cox, S. M. Kauzlarich and H. Hope for single crystal x-ray diffraction measurements, and H. Yao, S. A. Kivelson, R. M. Martin, S. C. Zhang, X. L. Qi and I. I. Mazin for discussions. ARPES experiments were performed at the Advance Light Source, which is operated by the Office of Basic Energy Science, U.S. Department of Energy. The Stanford work is supported by DOE Office of Basic Energy Science, Division of Materials Science and Engineering, under contract DE-AC02-76SF00515. Work at ORNL was supported by the DOE, Division of Materials Sciences and Engineering. MY thanks NSF Graduate Research Fellowship for financial support.

**Fig. 1.** Comparison between angle-integrated photoemission spectrum and calculated density of states. (a) and (c) show the valence band spectrum of LaOFeP and LaOFeAs, respectively. The inset shows the valence band of LSCO for comparison. (b) and (d) display the corresponding LDA density of states and projections onto the LAPW spheres.

**Fig. 2.** Fermi surface maps of LaOFeP and LaOFeAs. (a) and (c) are the unsymmetrized raw data from LaOFeP and LaOFeAs, respectively. Both maps are obtained by integrating the EDC's over an energy window of $E_F \pm 15$ meV. The red square highlights the boundary of the 1$^{st}$ zone. Note that we use the Brillouin zone corresponding to the two-iron unit cell with the M point at $(\pi, \pi)$, which is $(\pi, 0)$ in the large Brillouin zone for a simple iron square lattice. (b) and (d) illustrate the symmetrized FS map obtained by flipping and rotating the raw data shown in left panel along the high symmetry lines to reflect the symmetry of the crystal structure and to partially remove the suppression of the spectral weight due to matrix element effect.

**Fig. 3.** ARPES spectra (image plots) along two high symmetry lines: Γ-X and Γ-M for LaOFeP and LaOFeAs, respectively. For LaOFeP, LDA band structures along the same high symmetry lines (red lines) are superimposed on top of ARPES spectra. For better comparison with experimental data, the LDA band structures are shifted up by ~0.11 eV and then renormalized by a factor of 2.2. Three bands crossing $E_F$ are labeled as $\Gamma_1$, $\Gamma_2$ and $M$, corresponding to two hole pockets centered at Γ and an electron pocket centered at M, respectively.

**Fig. 4.** Energy distribution curves along the same high symmetry lines shown in Fig. 3. EDC's right at the $E_F$ crossing are highlighted in red. The leading-edge midpoints of these highlighted EDC's clearly reach $E_F$ for all bands crossing $E_F$, indicating, within our experimental uncertainty, the absence of a pseudogap in this system.

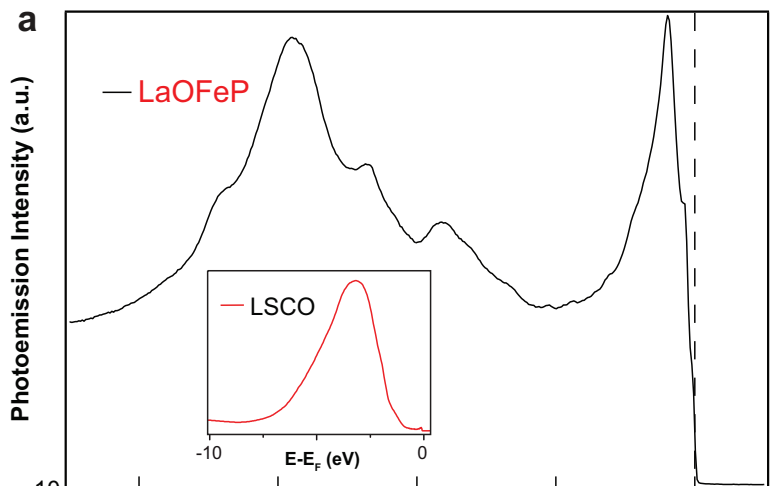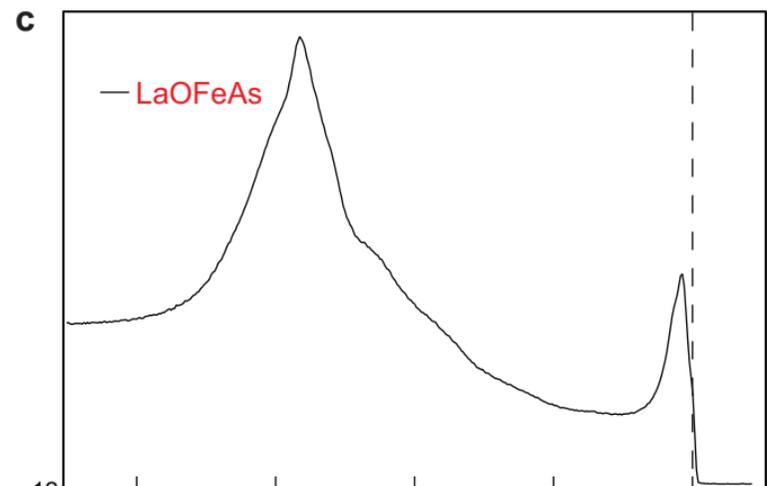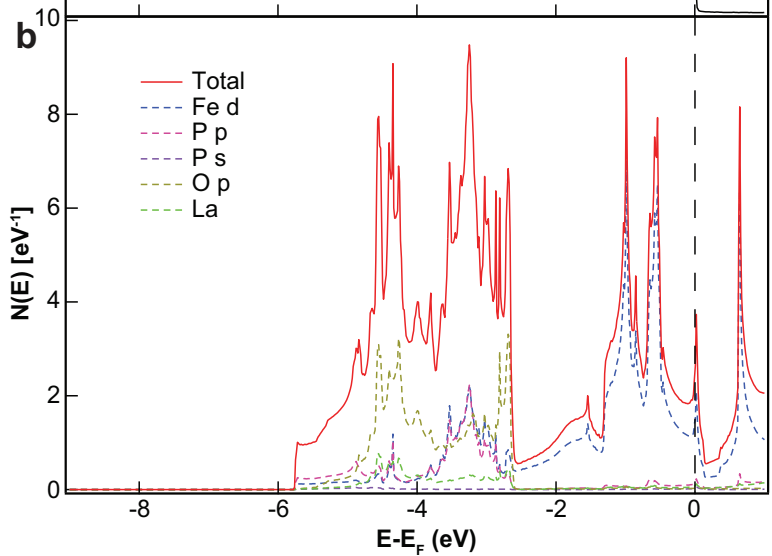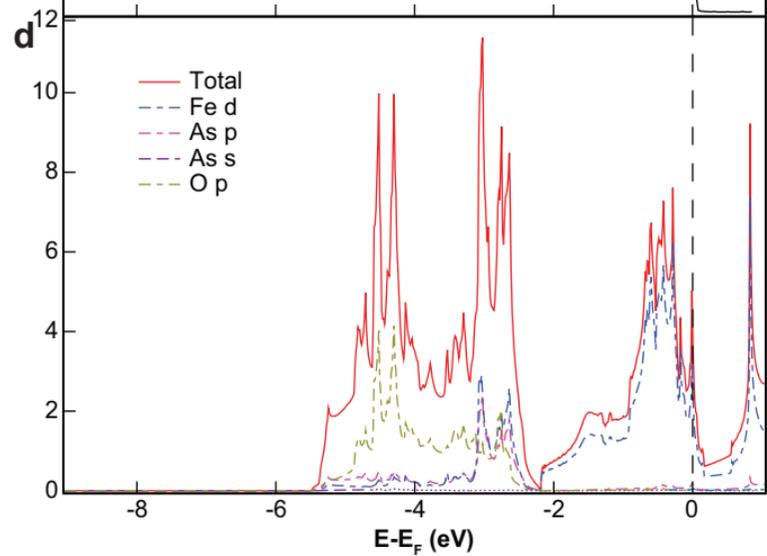

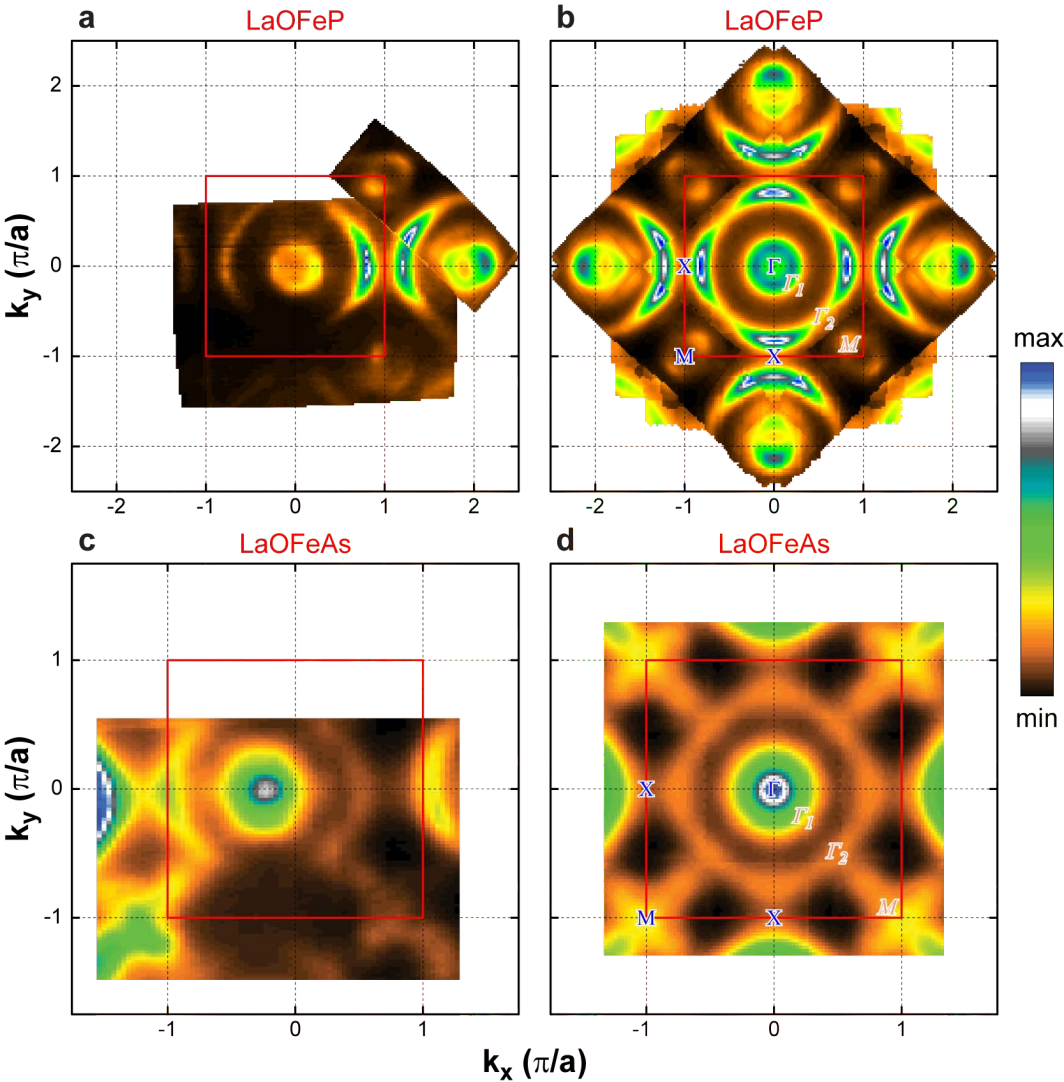

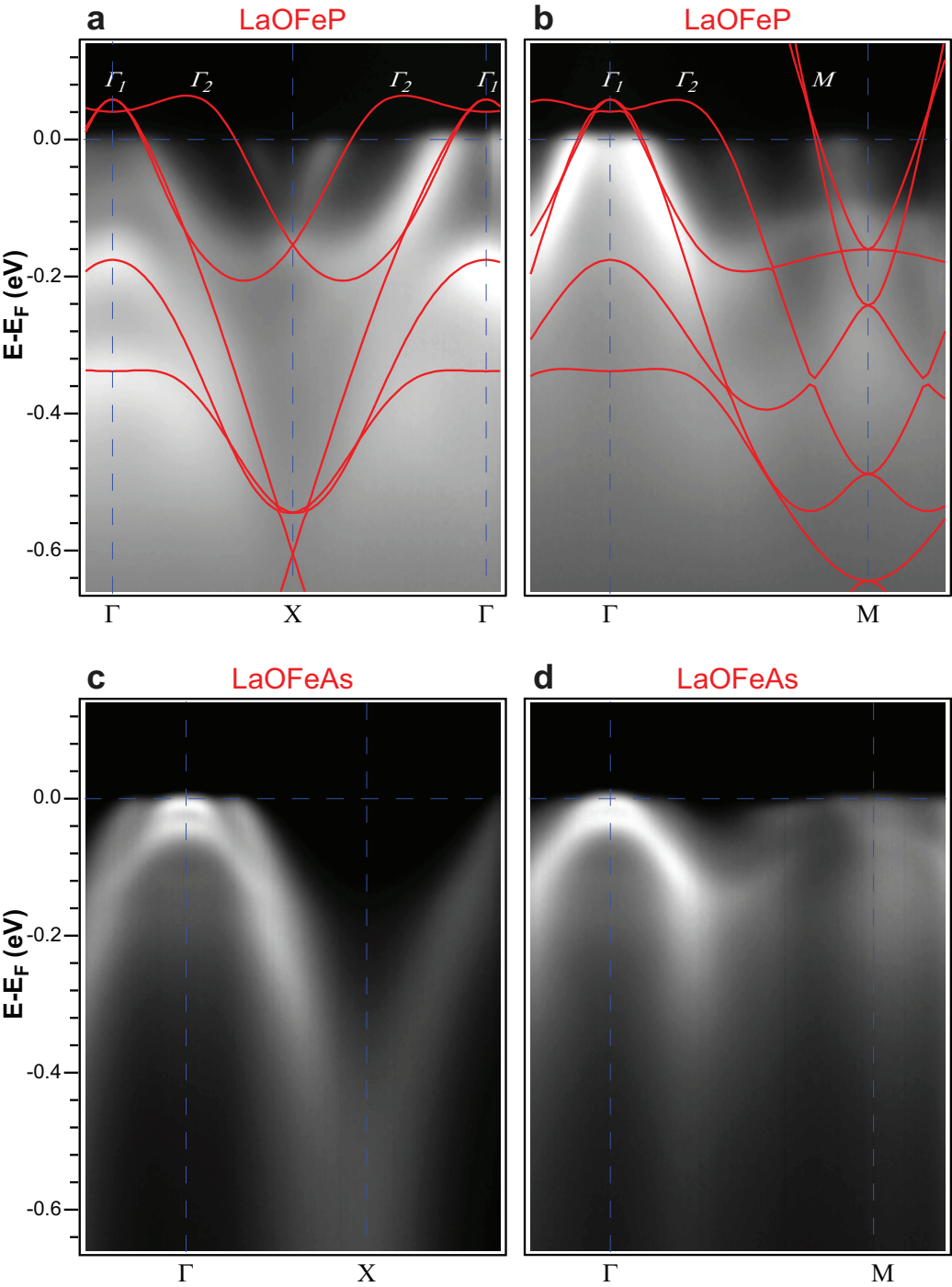

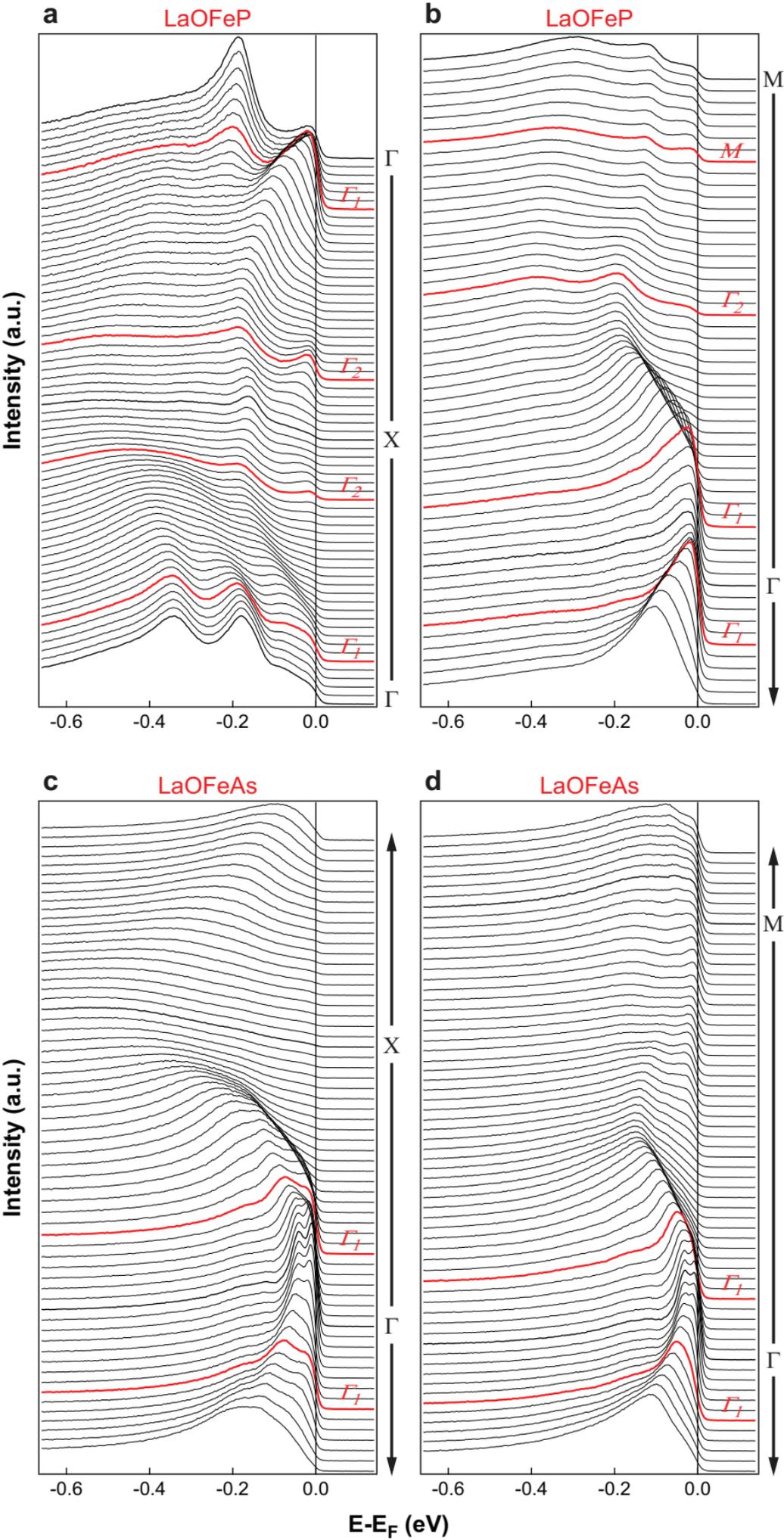